# Empirical study of performance of data binding in ASP.NET web applications

Toni Stojanovski[1], Marko Vučković[1], and Ivan Velinov[1]

[1]Faculty of Informatics, European University, Skopje, Republic of Macedonia,
toni.stojanovski@eurm.edu.mk, {vuckovic.marko, velinov.ivan}@live.eurm.edu.mk;

*Abstract*—Most developers use default properties of ASP.NET server controls when developing web applications. ASP.NET web applications typically employ server controls to provide dynamic web pages, and data-bound server controls to display and maintain database data. Though the default properties allow for fast creation of workable applications, creating a high-performance, multi-user, and scalable web application requires careful configuring of server controls and their enhancement using custom-made code. In providing commonly required functionality in data-driven ASP.NET web applications such as paging, sorting and filtering, our empirical study evaluated the impact of various technical approaches: automatic data binding in web server controls; data paging and sorting on web server; paging and sorting on database server; indexed and non-indexed database columns; clustered vs. non-clustered indices. The study observed significant performance differences between various technical approaches.

*Index terms* — web applications, scalability, database access

## 1. INTRODUCTION

In the last few years we are observing increased use of web applications. This is a consequence of many factors: zero-client installation, server-only deployment, powerful development tools, growing user base etc. Furthermore, competition and the quickly changing and growing user requirements create a demand for rapid development of web applications. Microsoft Visual Studio (MVS) is the dominant web applications development environment of today. MVS provides numerous mechanisms to support rapid development of ASP.NET applications. Most developers tend to use the default ASP.NET mechanisms: page caching; introduction of state in HTTP (session, cookies, hidden HTML controls etc.), data management, and the ASP.NET server controls which are arguably the most significant enabler of the rapid development. Though these mechanisms and ASP.NET server controls can significantly decrease the application's "time to market", at the same time they can reduce performance and scalability of the web application. Analysis of factors which influence the response time of web applications is an active area of research [1]. In this paper, we demonstrate the importance of adding custom program logics to ASP.NET server controls in order to improve performance and scalability of web applications. Here we put emphasis on the data binding mechanism, that is, the mechanisms used to maintain and display data. The other mechanism, such as data updating, page caching, data caching, state management etc. are left for future work.

Here we address the following research questions:
• What is the impact of the paging mechanism on the response time?
• What is the impact of indices on response time when sorting and paging the results?
• What is the dependence of the response time on the number of database records?
• What are the scenarios when it is better to use ASP.NET server controls? When is it better to use custom stored procedures for fetching, sorting and paging the results?

The outline of our paper is as follows. In Section 2 we are explaining the basics of data binding in ASP.NET applications, how paging is used to cut the expenses for fetching and displaying data, and sorting the data by some field. In Section 3 we are explaining our test environment and the testing approach. Test environment is sued to measure the response time of various ASP.NET pages which implement various methods for data fetching and display. In Section 4 we explain the results from the tests. The Section 5 concludes the paper and outlines further research.

2. DATA BINDING IN ASP.NET APPLICATION

When using ASP.NET data-bound control like GridView to display the data from a database, the fastest way is to bind the data-bound control with a data-source control, which connects to the database and executes the queries. When using this scenario, the data-source control automatically gets data from the database server [2] and displays it in the data-bound control. Data-source control gets the data from the database server after the Page.PreRender event in the page life cycle [3].

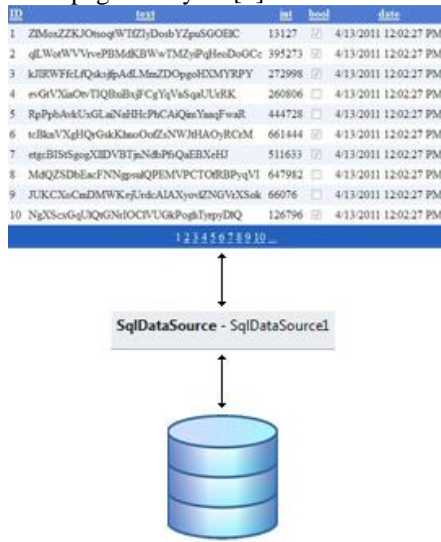

**Figure 1. Communication between a data-bound control and a database through a data-source control**

This is the code that is used for the data-source control to bind with the database.
```
<asp:SqlDataSource ID="SqlDS1" runat="server"
ConnectionString="<%$ ConnectionStrings:tdbConn %>"
SelectCommand="usp_autoDataBinding"
SelectCommandType="StoredProcedure"/>
```
Following code connects a GridView control with the data-source control.
```
<asp:GridView ID="GridView1" DataSourceID="SqlDS1" ...>
<Columns>
<asp:BoundField DataField="ID" HeaderText="ID"
SortExpression="ID" .../>
...
```

Another approach to display the data in a data-bound control is to get the data in the Page Load event, store it in a dataset object, and then bind the data-bound control to the dataset. We do not expect significant differences in execution time between the two scenarios, since the reasons for the slow response times (significant amounts of transmitted data, no use of database indices etc.) exist in both scenarios.

Following code shows how the GridView control is populated in the Page Load method.
```
SqlConnection connection = new SqlConnection(connString);
SqlCommand cmd = new SqlCommand("usp_autoDataBinding", connection);
cmd.Connection = connection;
cmd.CommandType = CommandType.StoredProcedure;
DataSet ds = new DataSet();
SqlDataAdapter sda = new SqlDataAdapter(cmd);
sda.Fill(ds);
DataView dv = new DataView(ds.Tables[0]);
dv.Sort = orderBy;
GridView1.PageIndex = pageNumber;
GridView1.DataSource = dv;
GridView1.DataBind();
```
The variables orderBy and pageNumber are taken from the query string (explained in Section 3).

Following stored procedure is used to query the data from the database
```
CREATE PROCEDURE [dbo].[usp_autoDataBinding]
AS
BEGIN
    SELECT * FROM testTable
END
```
**Code 1. Query that returns all data from a database.**

When there are a lot of records to display in a web page, it is a common practice to show only a limited number of records (a page of records) and to allow the user to navigate through the pages of records i.e. to use "data paging". Data-bound controls such as GridView can use the automatically provided mechanisms for sorting and paging in data-bound and data-source controls [2]. First, the data-source control gets **all** the data from the database (see Code 1), and then the ASP.NET data-bound control is responsible to sort the dataset and display only a small number of records enough to fill a page. For example, a dataset can contain millions of records, and a web page displays only 10 of these records. This approach poses two problems: (i) lots of data is transferred between the database server and the web server (in a multi-server deployment scenario which is dominant in the production environment); (ii) there is significant consumption of CPU and memory resources to sort large datasets. Clearly, these problems have significant negative impact on the performance and scalability of the application.

The impact of these problems can be reduced by decreasing the amount of data sent through the network, and reducing the consumption of resources. One needs to write a custom SQL stored procedure which sorts and returns only the records which will be displayed in the web page. Thus, the network consumption is reduced, and the database server gets the responsibility to sort and page the records. There are many ways to implement a stored procedure that can page and sort the results. We are using the following one:
```
CREATE PROCEDURE [dbo].[usp_selectGridViewOrderByID]
@pageNumber int,
@PageSize int = 10
AS
DECLARE @Ignore int
DECLARE @LastID int
```

```
IF @pageNumber > 1
BEGIN
    SET @Ignore = @PageSize * @pageNumber
    SET ROWCOUNT @Ignore
    SELECT @LastID = ID from testTable ORDER BY ID ASC
END
ELSE
BEGIN
    SET ROWCOUNT @PageSize
    SET @LastID = 0
END
SET ROWCOUNT @PageSize
SELECT * FROM testTable WHERE ID > @LastID ORDER BY ID ASC
```
**Code 2. SQL Stored procedure which supports custom data sorting and paging.**

This stored procedure logically divides the records from table testTable into pages of size @pageSize records, and returns the records from page @pageNumber. Records are ordered by field ID. Performance of this stored procedure greatly depends on the use of index on field ID and the type of index: clustered or non-clustered [4]. By using indexed data structure we can significantly improve the time required for getting information out of the database.

We expect major differences in response time depending on the following parameters:
- Number of records in database
- When paging and sorting is done by ASP.NET, or in SQL stored procedures
- Database indices
- Different deployment scenarios.

These scenarios based on the previous parameters will be tested in our test environment.

## 3. TESTING APPROACH

For our test environment we used HP 550 Notebook with following characteristics:
- Processor: Intel(R) Core (TM)2 Duo CPU T5470 @1.60 GHz
- RAM: 2.00 GB
- OS: Windows 7 Professional 32 – bit
- Internet Information Services (IIS) Version 7.5.7600.16385
- Visual Studio 2010 Ultimate
- SQL Server 2008 Express (use only 1GB RAM)

For the test environment we created a web application with three web pages – one for each of the data binding, paging and sorting approaches as explained in Section 2. Test database, named testdatabase, has one table named testTable. The database has five fields. Records in the table are populated with random values.

| Name | Type |
|---|---|
| ID | int, autoincrement |
| TextField | varchar(50) |
| IntField | int |
| BoolField | bit |
| DateField | datetime |

Web pages use a GridView control to display the data returned from the database. Each web page uses a different mechanism to bind the GridView control with the data, and to sort and page the data. Sorting field and page number are passed to the web page in the query string for the HTTP request.

We also created a windows application that sends HTTP request to the IIS web server. Query string for the HTTP request contains a randomly chosen page number, and the sorting field.

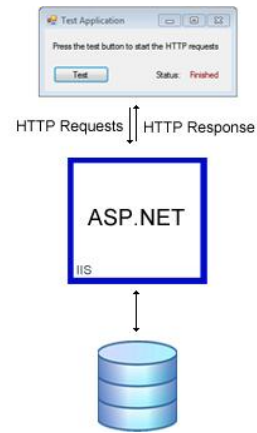

**Figure 2. HTTP requests in the test environment**

We are interested in the time required to process a HTTP request on the web server. Requested page queries data from a SQL Server table. We used the ASP.NET tracing to determine when certain events in the page lifecycle occur, that is, the start and end point in the page processing. We start the timer at Page_Init event and end the timer at Page_SaveStateComplete event, which is after the Page_PreRender event. As the requests are sent to ASP.NET, the page gets the variables from the query string and therefore chooses which stored procedure to use. Web pages are responsible to record the response time in a text file, and these time measurements are later analyzed.

The first page uses *Automatic Data Binding* (ADB). Page contains a GridView control, with paging and sorting allowed. The control is populated with a stored procedure that gets all the records from the database as in Code 1.

The second page populates a GridView control with the same stored procedure called by the first page, but this time we populate the GridView control in the Page_Load event handler instead of after the Page_preRender event.

The third page uses a custom stored procedure (see Code 2) to query the results. The stored procedure orders the results at the SQL server, and returns only the records that will be shown in the web page.

## 4. MAIN RESULTS

In Figure 3- Figure 5 we show the results when the data table has 1.000.000 records. Every figure represents one of the web pages mentioned earlier, and in every page there are three types of results, depending on which field was used to sort the results. At this time, table testTable has no indices. Every web page and the corresponding sorting and paging approach was tested 500 times. Measured response times were grouped into a small number of bins, usually 10. Figures given in this section show the frequency of the bins.

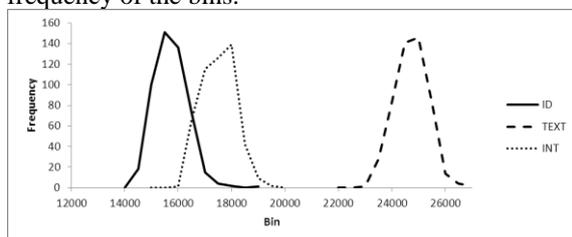

Figure 3. Automatic Data Binding.

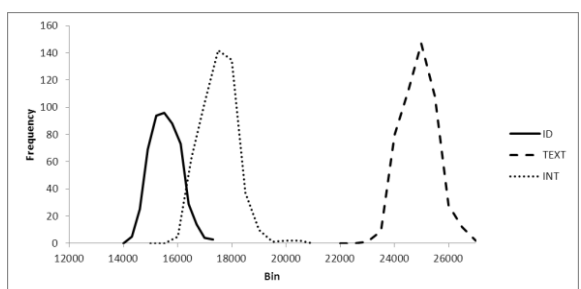

Figure 4. Data binding in Page_Load.

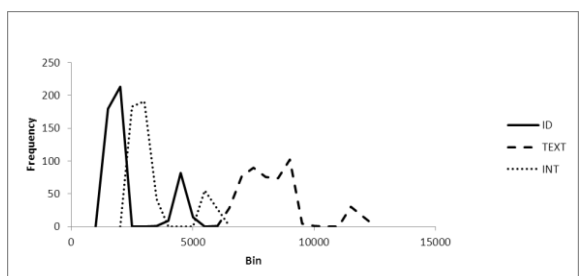

Figure 5. Custom Paging without indices.

Figure 3 and Figure 4 show the results when the data sorting and paging are executed in the ASP.NET web page using automatic data binding and filling the data in the data-bound control in the Page_Load method, respectively. Response times are very similar, as expected. The problem with this method is that the data-source control needs to fetch 1.000.000 results from the database before the ASP.NET can sort and page the results. At this time the records are not coming in predefined sort order, so the ASP.NET needs to check every record from the sort column before the final result set can be sorted as needed. Field ID is auto incremented, and the records are physically sorted by this field in the database. Therefore, the time needed for the ASP.NET to sort the result set by ID field is faster than the other two sorts. Response time when ordering by TextField and IntField is different because it is faster to sort integer than textual fields.

Figure 5 represents the results when the sorting and paging are done in the database server using an SQL stored procedure as in Code 2. The reader should note that Figure 5 uses a different time scale from Figure 3 and Figure 4. Response time is significantly shorter compared with the previous two approaches. The reason is twofold: (i) SQL server is optimized for working with large datasets; (ii) SQL stored procedure returns to the ASP.NET web page only a small number of records sufficient to fill a web page. The difference in the response time when sorting by different columns is caused by same reasons as explained for Figure 3 and Figure 4.

Next, we repeated the above tests when there are indices in the table testTable. The aim is to see the differences in response time when clustered and non-clustered indices are used.

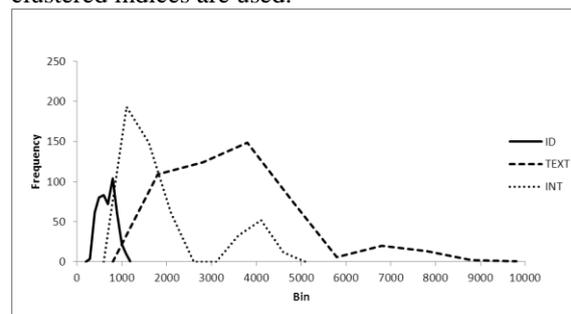

Figure 6. Custom Paging. Clustered index on ID.

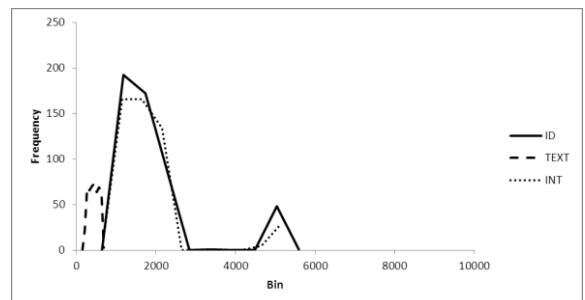

Figure 7. Custom Paging. Clustered index on TextField.

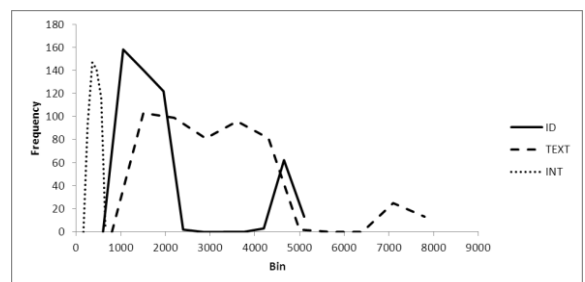

Figure 8. Custom Paging. Clustered index on IntField.

In Figure 6 the table testTable is clustered by ID, and fields IntField and TextField have non-clustered indices. Because of the clustered index, the response time when sorting by ID is significantly smaller compared to the response time when sorting by IntField and TextField. This pattern repeats in Figure 7 and Figure 8: the response time when sorting by a clustered index is shorter that the response time when sorting by a non-clustered index.

Comparing Figure 6, Figure 7 and Figure 8 with Figure 5 it becomes obvious that using either clustered or non-clustered indices provides significant improvements in the time required for data fetching. The presence of indices has no impact on the response time when the sorting and paging is done in the ASP.NET web page on the web server using the SQL stored procedure from Code 1. Results are identical to those shown in Figure 3 and Figure 4.

A peculiar property of Figures 5-8 is the presence of two peaks. They appear when sorting is done on a non-indexed field (all curves in Figure 5), or a field with a non-clustered index (IntField and TextField in Figure 6, ID and IntField in Figure 7, ID and TextField in Figure 8). This means that there are two different groups of time responses for the SQL stored procedure in Code 2. The problem lies in the second select statement "SELECT * FROM testTable WHERE ID > @LastID ORDER BY ID ASC" in Code 2. We detected that the response time is much longer when the input parameter @pageNumber < 18000. @LastID is smaller for smaller values of @pageNumber. Consequently, the SELECT statement sorts and returns a larger data set, and the time needed for its execution increases. The SQL server uses the index file to identify the ordering of records, and then joins the index file with the records from the table testTable, and finally returns every column in the record (note the use of the * sign which means that all table columns are returned). As the size of the recordset increases, the SQL server uses more memory to sort the recordset. If the recordset consumes more memory than what is available to the SQL server process, then the SQL server starts to use virtual memory which is much slower than the RAM. In our test environment, SQL server starts using the virtual memory when the number of records in the recordset is larger than 820,000 (first 18,000 pages with 10 records each are skipped). The above argument holds for both cases - sorting is done on a non-indexed field or a field with a non-clustered index.

However, when the sorting is done by the clustered index field (ID in Figure 6, TextField in Figure 7, IntField in Figure 8), then it can be noticed that there is a single peak. Records in the table are already physically ordered by the clustered index field.

We repeated the measurements for a different number of records in the table: 100.000, 200.000, 500.000, and 1.000.000 records. The aim was to test the dependency of the response time on the number of records. As expected, the response time is larger for larger number of database records. Response time grows faster with the number of records for ASP.NET server sorting and paging compared to SQL server sorting and paging. Figure 9- Figure 12 show the relation between response times averaged over 500 tests and the number of records in the table testTable. The fastest response time and the slowest growth with the number of records in the table testTable is achieved when using an SQL stored procedure with a clustered index, followed by an SQL stored procedure with a non-clustered index, followed by an SQL stored procedure without index, followed by web server sorting and paging.

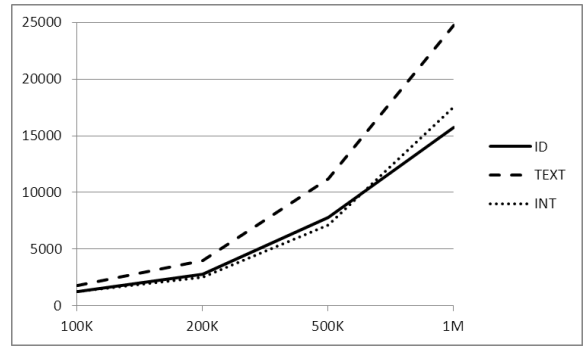

Figure 9 Average response time vs. number of table records when using Automatic Data Binding.

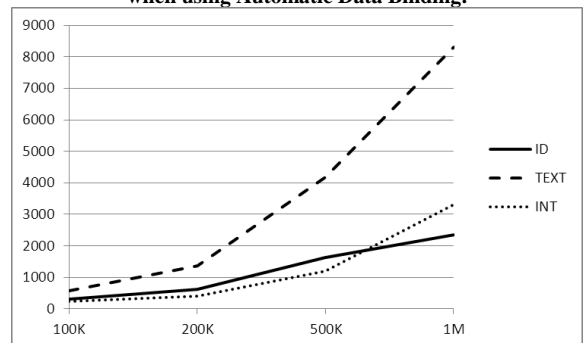

Figure 10 Average response time vs number of table records when using SQL stored procedure and no index.

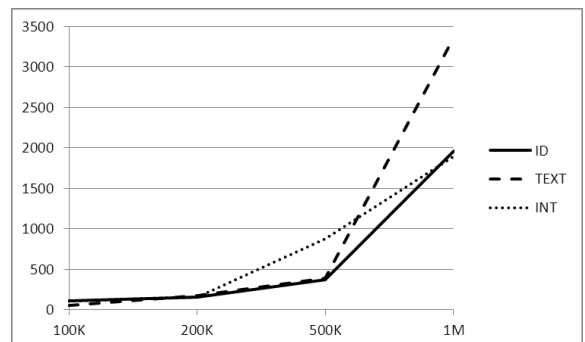

Figure 11 Average response time vs number of table records when using SQL stored procedure and sorting by non-clustered index.

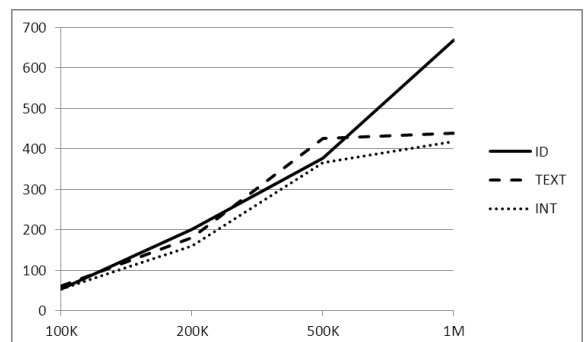

Figure 12 Average response time vs number of table records when using SQL stored procedure and sorting by clustered index.

Other relations between the response times as shown in Figure 3-Figure 5 are valid irrespective of the number of records have been preserved. The only difference worth of mentioning is shown in Figure 13. When the number of records in the table is smaller (e.g. 500,000) and the clustered index is on TextField, then there is only one group of response times when sorting by ID and IntField since the memory consumption is small enough not to trigger usage of virtual memory. Furthermore, the response time when sorting by ID and IntField and 500,000 records is significantly smaller compared to Figure 7 for 1,000,000 records.

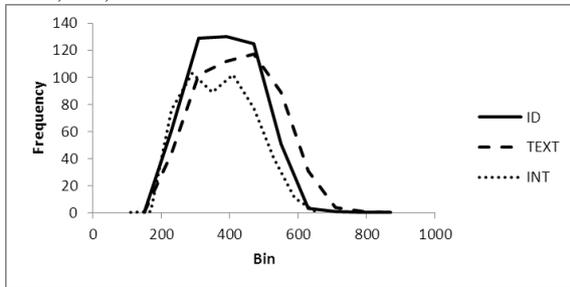

Figure 13. Custom Paging. Clustered index on TextField. 500,000 records in table testTable.

Above mentioned tests were repeated in a distributed deployment scenario: MS SQL server was installed on one physical server, and IIS web server was installed on another server. We have observed dramatic increases in the response time depending on the network speed for the test cases where the sorting and paging is done on the web server and consequently significant amounts of data travel over the network.

4.1 Improved SQL sorting and paging

The problem with the second select statement from Code 2 mentioned before in Section 4 can be solved by this modification of the stored procedure:
CREATE TABLE #t(x INT)
SET ROWCOUNT @PageSize
INSERT INTO #t
SELECT [int] FROM testTable WHERE [int] > @LastID ORDER BY [int] ASC

SELECT testTable.* FROM #t
LEFT JOIN testTable ON #t.[x] = testTable.[int]

**Code 3. Modifications to the SQL Stored procedure from Code 2.**

"SELECT *" statement from Code 2 is broken into two parts: First part orders the records by the indexed field and stores only the indexed field into a temporary table #t. Only @PageSize records are stored. No join is done between the index and the records in the table testTable, and thus the execution time is very short for the first part. Second part joins the records from the temporary table with the records from the original table testTable. Since the join is done on @PageSize records only (e.g. 10 records), the second part finishes very quickly too. Figure 14 demonstrates orders of magnitude improvement when the modified SQL stored procedure from Code 3 is used. Similar improvement is achieved when the sorting and paging is done on a field with non-clustered index.

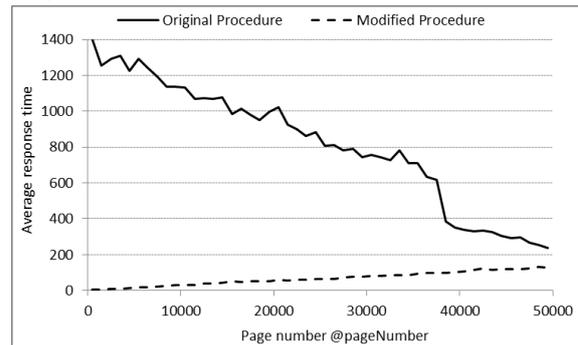

Figure 14. Average response time vs page number @pageNumber when sorting by clustered indices

5. CONCLUSION

Using the default options of the ASP.NET data bound controls allows for rapid development of sorting and paging functionality. If an ASP.NET data-source control is used to fetch all the data from the database, and then a data-bound control sorts and pages the recordset, then the response time can grow quickly with the size of the returned recordset.

An SQL stored procedures implementing sorting and paging on the SQL server ought to be used when high performance and low consumption of resources are required. It takes less time to fetch the recordset, and then to send to ASP.NET only the records that will be displayed. The response time can be further decreased if the sorting and paging is done on field with indices. Best results are achieved for clustered indices.